\documentclass[12pt]{article}
\usepackage{graphicx,amsmath,cite,subfig}
\usepackage{authblk}
\usepackage{geometry}[1]

\begin{document}
\title{Partial Integrability of 3-d Bohmian Trajectories}

\author{G. Contopoulos \thanks{gcontop@academyofathens.gr,}, A.C. Tzemos \thanks{thanasistzemos@gmail.com} and C. Efthymiopoulos \thanks{cefthim@academyofathens.gr}}

\affil{\small{Research Center for Astronomy and Applied Mathematics \\of the Academy of Athens \\ Soranou Efessiou 4, GR-11527 Athens, Greece}}

\date{}

\vspace{10pt}

\maketitle
\begin{abstract}
In this paper we study the integrability of 3-d Bohmian trajectories of a system of quantum harmonic oscillators. We show that the initial choice of quantum numbers is responsible for the existence (or not) of an integral of motion which confines the trajectories on certain invariant surfaces. We give a few examples of orbits in cases where there is or there is not an integral and make some comments on the impact of partial integrability in Bohmian Mechanics. Finally, we make a connection between our present results for the integrability in the 3-d case and analogous results found in   the 2-d and 4-d cases.
\end{abstract}

\section{Introduction}

Bohmian Mechanics is an alternative pilot wave interpretation of Quantum Mechanics, developed initially by Louis De Broglie and later by David Bohm \cite{debroglie1927a,debroglie1927b,Bohm, BohmII}. It has attracted the attention of many researchers  as a conceptual framework which provides insight to the fundamental notions of Quantum Mechanics \cite{holland1995quantum,deurr2009bohmian}. In recent years extensive studies have been carried out also on practical applications of Bohmian Mechanics in Atomic and Molecular Physics \cite{benseny, pladevall2012applied}.

A topic of particular interest regards the appearance of order and chaos in Bohmian Mechanics \cite{frisk1997properties, falsaperla2003motion, wisniacki2005motion, efthymiopoulos2006chaos, wisniacki2007vortex,efthymiopoulos2007nodal, contopoulos2008ordered, PhysRevE.79.036203, contopoulos2012order, Tzemos2016}. In general order and chaos coexist in a given system. In \cite{contopoulos2008ordered} we found that ordered orbits can be given approximately by formal series expansions. On the other hand it was found that chaos manifests itself when a Bohmian trajectory passes close to the moving nodal points of the wavefunction, where $\Psi=0$ \cite{wisniacki2005motion, efthymiopoulos2006chaos, wisniacki2007vortex,  efthymiopoulos2007nodal,PhysRevE.79.036203}. Orbits that never approach the nodal points are ordered \cite{contopoulos2012order}.

In \cite{efthymiopoulos2007nodal,PhysRevE.79.036203} we studied the onset of chaos in 2-d Bohmian systems. We found that chaos is introduced in the broad neighbourhood of the moving nodal points. In particular, chaos is not introduced very close to a nodal point, but when a trajectory approaches an unstable point (called X-point) at a certain distance from the nodal point. X-points are defined as the points where the velocity of the trajectory is equal to that of its corresponding nodal point. Nodal points and X-points form moving `nodal point-X-point complexes'. The nodal point-X-point mechanism of chaos was shown in \cite{PhysRevE.79.036203} to be generic in the 2-d case, i.e. it holds for arbitrary wavefunctions.

Extending this study to 3-d systems \cite{Tzemos2016}, we found the nodal points and their associated X-points in simple cases of 3-d harmonic oscillator quantum states, namely when 
$\Psi(t)=\frac{1}{\sqrt{3}}\Big(\Psi_{1,0,0}(t)+\Psi_{0,1,0}(t)+a_3\Psi_{0,0,1}(t)+a_4\Psi_{0,0,2}(t)\Big),$
where 
\begin{align}
\Psi_{n_1,n_2,n_3}(t)=\Psi_{n_1,n_2,n_3}(\vec{x})e^{-iE_it/\hbar}
\end{align}
and $\Psi_{n_1,n_2,n_3}(\vec{x})$ are eigenstates of the 3-d harmonic oscillator of the form
\begin{align}\label{eigenstate}
\Psi_{n_1,n_2,n_3}(\vec{x})=\prod_{k=1}^3\frac{\Big(\frac{m_k\omega_k}{\hbar\pi}\Big)^{\frac{1}{4}}\exp\Big(\frac{-m_k\omega_kx_k^2}{2\hbar}\Big)}{\sqrt{2^{n_k}n_k!}}H_{n_k}\Big(\sqrt{\frac{m_k\omega_k}\hbar}x_k\Big),
\end{align}
where $n_1,n_2,n_3$ stand for their quantum numbers, $\omega_1,\omega_2,\omega_3$ for their frequencies and $E_1,E_2,E_3$ for their energies. For a given set $(n_1,n_2,n_3)$ we have
\begin{align}
E=\sum_{i=1}^3(n_i+\frac{1}{2})\hbar\omega_i
\end{align}
In \cite{Tzemos2016} we found a particular case where an exact (analytic) integral of motion exists. This integral confines Bohmian trajectories on certaint invariant surfaces. 
In particular, when $a_4=0$ the orbits lie on spherical surfaces
$x_1^2+x_2^2+x_3^2=C$.
Both ordered and chaotic orbits lie on particular spherical surfaces and the system is partially integrable.
On the other hand for small values of $a_4$ we observed 3-d diffusion of the trajectories inwards and outwards the spherical surfaces and showed that the ``3-d structures of nodal and X-points'' of the unperturbed case ($a_4=0$) could still describe qualitatively the generation of chaos. 

In the present paper we find general cases where an integral of motion in 3-d configuration space exists for a system of harmonic oscillators. In these cases the motion takes place on integral surfaces. Thus the problem becomes essentially 2 dimensional. However in general there is no further integral of motion, consequently we may have both order and chaos on the same 2-d  integral surface.

One relevant physical consequence of partial integrability refers to the phenomenon of `quantum relaxation' \cite{VALENTINI19915, VALENTINI19911, valentini2005dynamical}, namely the posibility to derive a dynamical origin of Born's rule $P=|\Psi|^2$ using Bohmian trajectories, even if the initial distribution of the Bohmian particles $P_0$ does not satisfy this rule. It has been established that chaos is a necessary condition for a dynamical approach of Bohmian trajectories to Born's rule \cite{efthymiopoulos2006chaos, contopoulos2012order}. In partially integrable systems, as in general 2-d systems, chaos is generated by the 'nodal point-X-point' mechanism. On the other hand, as no diffusion is permitted outside every integral surface, partial integrability restricts quantum relaxation. Consequently, our present results can be viewed as examples of a general class of quantum states in which quantum relaxation \textit{cannot} be fully accomplished. In fact as shown in \cite{Tzemos2016}, the addition of a small perturbation to a partially integrable system results in a quite slow chaotic diffusion outside the integral surface. The subject of the rate of quantum relaxation in this case is proposed for further study.

In the sequel we present a systematic analysis of the general case of wavefunctions of the form
\begin{align}\label{general}
\Psi(t)=&a\Psi_{p_1,p_2,p_3}(t)+b\Psi_{r_1,r_2,r_3}(t)+c\Psi_{s_1,s_2,s_3}(t),
\end{align}
with $p_i, r_i, s_i,\,\, i=1,2,3$ arbitrarily chosen integers.
In section 2 we present a number of partially integrable cases with wavefunctions of the form (\ref{general}). In section 3 we study various special cases.
In section 4 we present numerical examples of ordered and chaotic behavior of Bohmian trajectories in the presence/absence of partial integrability. In section 5 we give some details for integrals in 2-d and 4-d cases and discuss their connection with the results of the previous sections. Finally, in section 6 we summarize our conclusions.

\section{Equations of Motion}

The Bohmian equations of motion in the 3-d case for an arbitrary wavefunction $\Psi=\Psi_R+i\Psi_{I}$ in a system of units such that $\hbar=1$ are:

\begin{align}\label{bohmeq}
m_i\frac{dx_i}{dt}=\Im\Big(\frac{\nabla_i \Psi}{\Psi}\Big)=\frac{1}{G}\Big(\frac{\partial \Psi_I}{\partial x_i}\Psi_R-\frac{\partial \Psi_R}{\partial x_i}\Psi_I\Big),\quad i=1,2,3
\end{align}
with $G=\Psi_R^2+\Psi_I^2$.
The wavefunction (\ref{general}) is now
\begin{align}
\nonumber\Psi(t)&=e^{-\frac{\sum_{i=1}^3m_i\omega_ix_i^2}{2}}\Big(aK_1H_{p1}H_{p2}H_{p3}e^{-iE_1t}
\\&+bK_2H_{r1}H_{r2}H_{r3}e^{-iE_2t}+cK_3H_{s1}H_{s2}H_{s3}e^{-iE_3t}\Big)
\end{align}
where $H_{pi},H_{ri},H_{si}$ are Hermite polynomials of $\sqrt{m_i\omega_i}x_i, i=1,2,3$. Moreover 
 $K_j=\frac{(m_1m_2m_3\omega_1\omega_2\omega_3)^{\frac{1}{4}}}{\pi^{\frac{3}{4}}\sqrt{2^{n_1+n_2+n_3}n_1!n_2!n_3!}}$ with $n_i=p_i$ if $j=1$, $n_i=r_i$ if $j=2$ and $n_i=s_i$ if $j=3$ (we assume that $E_1, E_2, E_3$ are different).
Without loss of generality we work with $m_1=m_2=m_3=1$. 
Using now the equations (\ref{bohmeq}) we find after some algebra:
\begin{align}\label{bohm1}
\nonumber\dot{x_1}=&\frac{1}{\tilde{G}}\Big(abK_1K_2H_{p_2}H_{p_3}H_{r_2}H_{r_3}[H_{p_1},H_{r_1}]\sin(\Delta E_{12}t)\\&\nonumber+acK_1K_3H_{p_2}H_{p_3}H_{s_2}H_{s_3}[H_{p_1},H_{s_1}]\sin(\Delta E_{13}t)\\&+bcK_2K_3H_{r_2}H_{r_3}H_{s_2}H_{s_3}[H_{r_1},H_{s_1}]\sin(\Delta E_{23}t)\Big)\\
\label{bohm2}\nonumber\dot{x_2}=&\frac{1}{\tilde{G}}\Big(abK_1K_2H_{p_1}H_{p_3}H_{r_1}H_{r_3}[H_{p_2},H_{r_2}]\sin(\Delta E_{12}t)\\&\nonumber+acK_1K_3H_{p_1}H_{p_3}H_{s_1}H_{s_3}[H_{p_2},H_{s_2}]\sin(\Delta E_{13}t)\\&+bcK_2K_3H_{r_1}H_{r_3}H_{s_1}H_{s_3}[H_{r_2},H_{s_2}]\sin(\Delta E_{23}t)\Big)\\
\label{bohm3}\nonumber\dot{x_3}=&\frac{1}{\tilde{G}}\Big(abK_1K_2H_{p_1}H_{p_2}H_{r_1}H_{r_2}[H_{p_3},H_{r_3}]\sin(\Delta E_{12}t)\\&\nonumber+acK_1K_3H_{p_1}H_{p_2}H_{s_1}H_{s_2}[H_{p_3},H_{s_3}]\sin(\Delta E_{13}t)\\&+bcK_2K_3H_{r_1}H_{r_2}H_{s_1}H_{s_2}[H_{r_3},H_{s_3}]\sin(\Delta E_{23}t)\Big)
\end{align}

In the above expressions we have set $[H_A,H_B]\equiv H_AH_B'-H_A'H_B$, $\tilde{G}=Ge^{\omega_1x_1^2+\omega_2x_2^2+\omega_3x_3^2}$ and $\Delta E_{ij}=E_i-E_j$. Furthermore $H_{p_i}'= \frac{dH_{p_i}(\sqrt{\omega_1}x_i)}{dx_i}$,  $H_{r_i}'=\frac{dH_{r_i}(\sqrt{\omega_2}x_i)}{dx_i}$ and $H_{s_i}'= \frac{dH_{s_i}(\sqrt{\omega_3}x_i)}{dx_i}$.

We now find a combination of Eqs.~(\ref{bohm1})-(\ref{bohm3}) of the form
\begin{align}\label{formd}
\dot{x}_1f_1(x_1)+\dot{x}_2f_2(x_2)+\dot{x}_3f_3(x_3)
\end{align}
that is equal to zero. Then the corresponding integral of motion is
\begin{align}\label{form}
\int f_1(x_1)dx_1+\int f_2(x_2)dx_2+\int f_3(x_3)dx_3=C.
\end{align}
We notice that in Eqs.~(\ref{bohm1})-(\ref{bohm3}) there are 3 different trigonometric terms multiplied by functions of all three variables $x_1, x_2, x_3$. Therefore it is not possible in general to find common factors $f_1(x_1),f_2(x_2),f_3(x_3)$ in order to eliminate all trigonometric terms. However if we impose certain restrictions on the integers $p,r,s$, we can eliminate all those trigonometric terms.  

In particular: we can eliminate the trigonometrig term $\sin(\Delta E_{12}t)$ in $\dot{x}_1$ by setting $p_1=r_1$  (thus $[H_{p_1},H_{r_1}]=0$), the  trigonometric term $\sin(\Delta E_{23}t)$ in $\dot{x}_2$ by setting $r_2=s_2$ (thus $[H_{r_2},H_{s_2}]=0$) and the trigonometric term $\sin(\Delta E_{13}t)$ in $\dot{x}_3$ by setting $s_3=p_3$ (thus $[H_{s_3},H_{p_3}]=0$). Then in $\dot{x}_1$ we have $[H_{r_1},H_{s_1}]=-[H_{s_1},H_{p_1}]$, in  $\dot{x}_2$ we have $[H_{p_2},H_{r_2}]=-[H_{s_2},H_{p_2}]$ and  in $\dot{x}_3$ we have $[H_{p_3},H_{r_3}]=-[H_{r_3},H_{s_3}]$. Therefore if we multiply $\dot{x}_1$ by $H_{r_1}H_{s_1}/[H_{r_1},H_{s_1}]$, $\dot{x}_2$ by $H_{p_2}H_{r_2}/[H_{p_2},H_{r_2}]$ and $\dot{x}_3$ by $H_{p_3}H_{r_3}/[H_{p_3},H_{r_3}]$ and add these terms we eliminate all three trigonometric terms. This yields:
\begin{itemize}
\item{Case $1$:  $p_1=r_1, r_2=s_2, s_3=p_3$, i.e.
$\Psi(t)=a\Psi_{p_1,p_2,p_3}(t)+b\Psi_{p_1,r_2,r_3}(t)+c\Psi_{s_1,r_2,p_3}(t).$
The conserved quantity is
\begin{align}\label{diff1}
\frac{\dot{x}_1H_{s_1}H_{p_1}}{[H_{s_1},H_{p_1}]}+\frac{\dot{x}_2H_{p_2}H_{r_2}}{[H_{p_2},H_{r_2}]}+\frac{\dot{x}_3H_{r_3}H_{p_3}}{[H_{r_3},H_{p_3}]}=0
\end{align}
and if we integrate these terms we find the integral
\begin{align}\label{olok}
\int\frac{\Big(\frac{H_{p_1}}{H_{s_1}}\Big)}{\Big(\frac{H_{p_1}}{H_{s_1}}\Big)'}dx_1+\int\frac{\Big(\frac{H_{r_2}}{H_{p_2}}\Big)}{\Big(\frac{H_{r_2}}{H_{p_2}}\Big)'}dx_2+\int\frac{\Big(\frac{H_{p_3}}{H_{r_3}}\Big)}{\Big(\frac{H_{p_3}}{H_{r_3}}\Big)'}dx_3=C.
\end{align}
Therefore the system is partially integrable and the orbits lie on surfaces given by Eq.~(\ref{olok}).}

\item{Case 2: Eliminate $\sin(\Delta E_{12}t)$ from $\dot{x}_1$, $\sin(\Delta E_{13}t)$ from $\dot{x}_2$ and $\sin(\Delta E_{23}t)$ from $\dot{x}_3$ by setting $r_1=p_1, s_2=p_2$ and $s_3=r_3$, i.e.
$\Psi(t)=a\Psi_{p_1,p_2,p_3}(t)+b\Psi_{p_1,r_2,r_3}(t)+c\Psi_{s_1,p_2,r_3}(t).$
Then we have the same relation (\ref{diff1})
and the corresponding integral is the one of Eq.~(\ref{olok}).
\item{Case 3: Eliminate $\sin(\Delta E_{23}t)$ from $\dot{x}_1$, $\sin(\Delta E_{12}t)$ from $\dot{x}_2$ and $\sin(\Delta E_{13}t)$ from $\dot{x}_3$ by setting $s_1=r_1, r_2=p_2$ and $s_3=p_3$, i.e. 
$\Psi(t)=a\Psi_{p_1,p_2,p_3}(t)+b\Psi_{r_1,p_2,r_3}(t)+c\Psi_{r_1,s_2,p_3}(t).$
Then we have 
\begin{align}\label{diff3}
\frac{\dot{x}_1H_{p_1}H_{r_1}}{[H_{p_1},H_{r_1}]}+\frac{\dot{x}_2H_{s_2}H_{p_2}}{[H_{s_2},H_{p_2}]}+\frac{\dot{x}_3H_{r_3}H_{p_3}}{[H_{r_3},H_{p_3}]}=0
\end{align} and 
\begin{align}
\int\frac{\Big(\frac{H_{r_1}}{H_{p_1}}\Big)}{\Big(\frac{H_{r_1}}{H_{p_1}}\Big)'}dx_1+\int\frac{\Big(\frac{H_{p_2}}{H_{s_2}}\Big)}{\Big(\frac{H_{p_2}}{H_{s_2}}\Big)'}dx_2+\int\frac{\Big(\frac{H_{p_3}}{H_{r_3}}\Big)}{\Big(\frac{H_{p_3}}{H_{r_3}}\Big)'}dx_3=C.
\end{align}
}
\item{Case 4: Eliminate $\sin(\Delta E_{23}t)$ from $\dot{x}_1$, $\sin(\Delta E_{13}t)$ from $\dot{x}_2$ and $\sin(\Delta E_{12}t)$ from $\dot{x}_3$ by setting $s_1=r_1, s_2=p_2$ and $r_3=p_3$, $\Psi(t)=a\Psi_{p_1,p_2,p_3}(t)+b\Psi_{r_1,r_2,p_3}(t)+c\Psi_{r_1,p_2,s_3}(t).$ Then 
\begin{align}\label{diff4}
\frac{\dot{x}_1H_{p_1}H_{r_1}}{[H_{r_1},H_{p_1}]}+\frac{\dot{x}_2H_{r_2}H_{p_2}}{[H_{r_2},H_{p_2}]}+\frac{\dot{x}_3H_{s_3}H_{p_3}}{[H_{s_3},H_{p_3}]}=0
\end{align} and 
\begin{align}
\int\frac{\Big(\frac{H_{r_1}}{H_{p_1}}\Big)}{\Big(\frac{H_{r_1}}{H_{p_1}}\Big)'}dx_1+\int\frac{\Big(\frac{H_{p_2}}{H_{r_2}}\Big)}{\Big(\frac{H_{p_2}}{H_{r_2}}\Big)'}dx_2+\int\frac{\Big(\frac{H_{p_3}}{H_{s_3}}\Big)}{\Big(\frac{H_{p_3}}{H_{s_3}}\Big)'}dx_3=C.
\end{align}}
\item{Case 5: Eliminate $\sin(\Delta E_{13}t)$ from $\dot{x}_1$, $\sin(\Delta E_{12}t)$ from $\dot{x}_2$ and $\sin(\Delta E_{23}t)$ from $\dot{x}_3$ by setting $s_1=p_1, r_2=p_2$ and $s_3=r_3$, i.e. 
$\Psi(t)=a\Psi_{p_1,p_2,p_3}(t)+b\Psi_{r_1,p_2,r_3}(t)+c\Psi_{p_1,s_2,r_3}(t)$.
Then we have the same relation (\ref{diff3}) and the integral is the same to that of Case 3.}
\item{Case 6: Eliminate $\sin(\Delta E_{12}t)$ from $\dot{x}_1$, $\sin(\Delta E_{13}t)$ from $\dot{x}_2$ and $\sin(\Delta E_{12}t)$ from $\dot{x}_3$ by setting $s_1=p_1, s_2=r_2$ and $r_3=p_3$, i.e. $\Psi(t)=a\Psi_{p_1,p_2,p_3}(t)+b\Psi_{r_1,r_2,p_3}(t)+c\Psi_{p_1,r_2,s_3}(t)$.
Then we have the same relation (\ref{diff4}) and the integral is the same to that of Case 4.}
}\end{itemize}

\section{Special Cases}
\begin{enumerate}
\item{Up to now we have assumed that no two of $E_1,E_2,E_3$ are equal. However in a resonant case we may have e.g. $E_1=E_2$. Then in Case 1 the terms with $\sin(\Delta E_{12}t)$ in $\dot{x}_1,\dot{x}_2,\dot{x}_3$ are zero and one may think that the restriction $p_1=r_1$ is not necessary. But we need to have $[H_{r_1},H_{s_1}]=-[H_{p_1},H_{s_1}]$ in order to be able to divide $\dot{x}_1$ by this quantity. Thererefore we must have $p_1=r_1$ besides the other two relations of Case $1$ ($r_2=s_2$ and $s_3=p_3$ ).
We conclude that the resonant cases have integrals of the same form as in the nonresonant ones. Only in the particular resonant case $E_1=E_2=E_3$ we have $\dot{x}_1=\dot{x}_2=\dot{x}_3=0$, i.e. the orbits become isolated points.}

\item{In all of the above cases we have assumed that $a,b,c\neq 0$. Otherwise if one factor, say $c$, is zero then we have only one trigonometric term $\sin(\Delta E_{12}t)$ and we find 
\begin{align}
\frac{\dot{x}_1H_{p_1}H_{r_1}}{[H_{p_1},H_{r_1}]}=\frac{\dot{x}_2H_{p_2}H_{r_2}}{[H_{p_2},H_{r_2}]}=\frac{\dot{x}_3H_{p_3}H_{r_3}}{[H_{p_3},H_{r_3}]}.
\end{align}
Then assuming that $p_1,p_2,p_3$ are not all equal to $r_1,r_2,r_3$ (if some indices are equal the corresponding terms are zero)
\begin{align}
\int\frac{\Big(\frac{H_{r_1}}{H_{p_1}}\Big)}{\Big(\frac{H_{r_1}}{H_{p_1}}\Big)'}dx_1=\int\frac{\Big(\frac{H_{r_2}}{H_{p_2}}\Big)}{\Big(\frac{H_{r_2}}{H_{p_2}}\Big)'}dx_2+C_1=\int\frac{\Big(\frac{H_{r_3}}{H_{p_3}}\Big)}{\Big(\frac{H_{r_3}}{H_{p_3}}\Big)'}dx_3+C_2.
\end{align}
Hence we have 2 integrals of motion.
In this case the problem is reduced to 1-d. The two integrals define two surfaces whose intersections are the orbits. No chaos appears at all in such cases.}

\item{Finally, in the cases 1-6 of the previous sections we have assumed only 3 inequalities between the 9 indices $p_i,r_i,s_i (i=1\dots 3)$ in Eq.~(\ref{general}). However if we have at the same time two cases, then we have 5 inequalities. E.g.  if we  combine cases $1$ and $2$, namely $s_1=r_1, s_2=p_2=r_2, s_3=p_3=r_3$, we find $K_2=K_3$ and $E_2=E_3$. Then the last two terms of the wavefunction, $\Psi_{r_1,r_2,r_3}(t)$ and $\Psi_{s_1,s_2,s_3}(t)$ become equal. Therefore we have 
\begin{align}
\Psi=a\Psi_{p_1,p_2,p_3}e^{-iE_1t}+(b+c)\Psi_{r_1,p_2,p_3}e^{-iE_2t}.
\end{align}
Therefore this case is like the special case 2. We have two integrals of motion and no chaos. If $E_1=E_2$ or $E_1=E_3$, we have similar results.}

\item{The case 
\begin{align}\Psi(t)=\Psi_{l,k,k}(t)+\Psi_{k,m,k}(t)+\Psi_{k,k,n}(t):
\end{align} In this case $p_1=l, p_2=p_3=k$ $r_1=r_3=k$, $r_2=m$ and $s_1=s_2=k$, $s_3=n$. Therefore this is a subcase of Case $4$. However in the integrals we have to distinguish the $k's$ in the first, second or third place ($k_1,k_2,k_3$ respectively), because they represent Hermite polynomials with respect to $\sqrt{\omega_1}x_1,\sqrt{\omega_2}x_2$ and $\sqrt{\omega_3}x_3$ correspondingly. Thus in this case we have the conserved quantity
\begin{align}
&\frac{\dot{x}_1H_lH_{k_1}}{[H_{l},H_{k_1}]}+\frac{\dot{x}_2H_mH_{k_2}}{[H_{m},H_{k_2}]}+\frac{\dot{x}_3H_nH_{k_3}}{[H_{n},H_{k_3}]}=0.
\end{align} }
\item{In the particular case that some indices are equal to zero, the corresponding Hermite polynomials are equal to 1 and their derivatives are zero. E.g. in the case
\begin{align}\label{sc}
\Psi=a\Psi_{l,0,0}(t)+b\Psi_{0,m,0}(t)+c\Psi_{0,0,n}(t)
\end{align}
we have $\frac{\dot{x}_1H_l}{H_l'}+\frac{\dot{x}_2H_m}{H_m'}+\frac{\dot{x}_3H_n}{H_n'}=0$
From the identity relations between Hermite polynomials we have 
$\frac{dH_k(z)}{dz}=2kH_{k-1}(z)\label{ident1}$
and
$H_k(z)=2zH_{k-1}(z)-\frac{dH_{k-1}(z)}{dz}\label{ident2}$.
The Hermite polynomials are functions of $\sqrt{\omega_i}x_i$. Consequently after some algebra we write the corresponding integral 
\begin{align}\label{integral}
\sum_{i=1}^3\frac{x_i^2-\frac{1}{\omega_i}\ln |H_{k_i-1}|}{k_i}=C,
\end{align} with $k_1=l, k_2=m, k_3=n$.
If $l=0$ or $m=0$ or $n=0$ then the corresponding terms in Eq.~(\ref{integral}) are zero and the integral depends only on $x_2, x_3$ or $x_1,x_3$ or $x_1, x_2$.}
\item{An even more special case is 
\begin{align}
\Psi(t)=a\Psi_{1,0,0}(t)+b\Psi_{0,1,0}(t)+c\Psi_{0,0,n}(t).
\end{align} In our previous paper \cite{Tzemos2016}  we had $n=1$ and found a precisely spherical integral surface $\sum_{i=1}^3x_i^2=R^2.$}
Similarly for $n=2$ we have
\begin{align}
x_1^2+x_2^2+\frac{x_3^2}{2}-\frac{\ln|x_3|}{2\omega_3}=C,
\end{align} and for $n=3$,
\begin{align}
x_1^2+x_2^2+\frac{x_3^2}{3}-\frac{\ln|4\omega_3x_3^2-2|}{3\omega_3}=C.
\end{align}
\end{enumerate}

\begin{figure}[ht]
\centering
\includegraphics[scale=0.4]{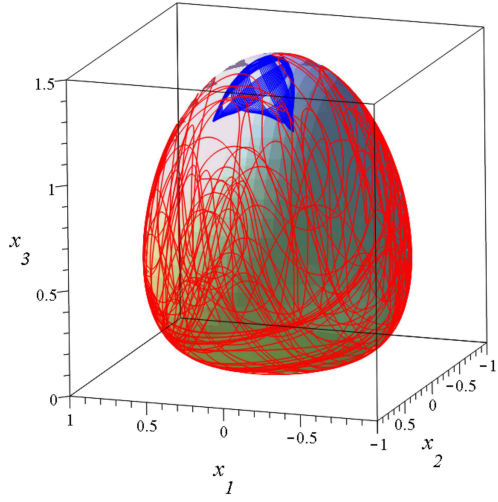}
\caption{Two orbits of the wavefunction $\Psi(t)=\frac{1}{\sqrt{3}}\Big(\Psi_{1,0,0}(t)+\Psi_{0,1,0}(t)+\Psi_{0,0,2}(t)\Big)$ on the surface $x_1^2+x_2^2+\frac{x_3^2}{2}-\frac{\sqrt{3}\ln|x_3|}{6}=2.392557011$ ($\omega_1=1,\omega_2=\sqrt{2},\omega_3=\sqrt{3}$). The blue one is ordered  ($x_1(1)=0.1647154159,x_2(1)=0.3,x_3(1)=1.4$) while the red one is chaotic ($x_1(1) = 0.6403124237, x_2(1) = 0.3,x_3(1)=1$).}\label{integral_surfaces_closed}
\end{figure}
In all previous cases $\Psi$ is the sum of three terms. On the other hand, if $\Psi$ is the sum of more than three terms we cannot find any combination of the indices that gives an integral of motion of the form (\ref{form}). 
However, we leave as an open problem whether there are (partially) integrable cases with different forms of integrals.

\section{Numerical Results}
We give now three numerical examples of partially integrable Bohmian trajectories, case A with a closed invariant surface, case B with an open invariant surface and case C with no invariant surface at all:
\begin{itemize}
\item{Case A: \begin{align}
\Psi(t)=\frac{1}{\sqrt{3}}\Big(\Psi_{1,0,0}(t)+\Psi_{0,1,0}(t)+\Psi_{0,0,2}(t)\Big).
\end{align}
The corresponding integral is $x_1^2+x_2^2+\frac{x_3^2}{2}-\frac{\ln|x_3|}{2\omega_3}=C$ and is represented by a closed surface. On this surface we have both ordered and chaotic orbits (Fig.~\ref{integral_surfaces_closed}).}
\item{Case B:\begin{align}\Psi(t)=\frac{1}{\sqrt{3}}\Big(\Psi_{0,0,0}(t)+\Psi_{1,1,0}(t)+\Psi_{1,0,2}(t)\Big)\end{align} This is a subcase of Case 1, because $p_1=p_2=p_3=r_3=s_2=0$ and $r_1=r_2=s_1=1$. Therefore in this case we have the integral $
-x_1^2+x_2^2+\frac{x_3^2}{2}-\frac{\ln|x_3|}{2\omega_3}=C$.

\begin{figure}
\subfloat[\label{integral_surfaces1}]
{\includegraphics[height=7.5cm, width=7.5cm]{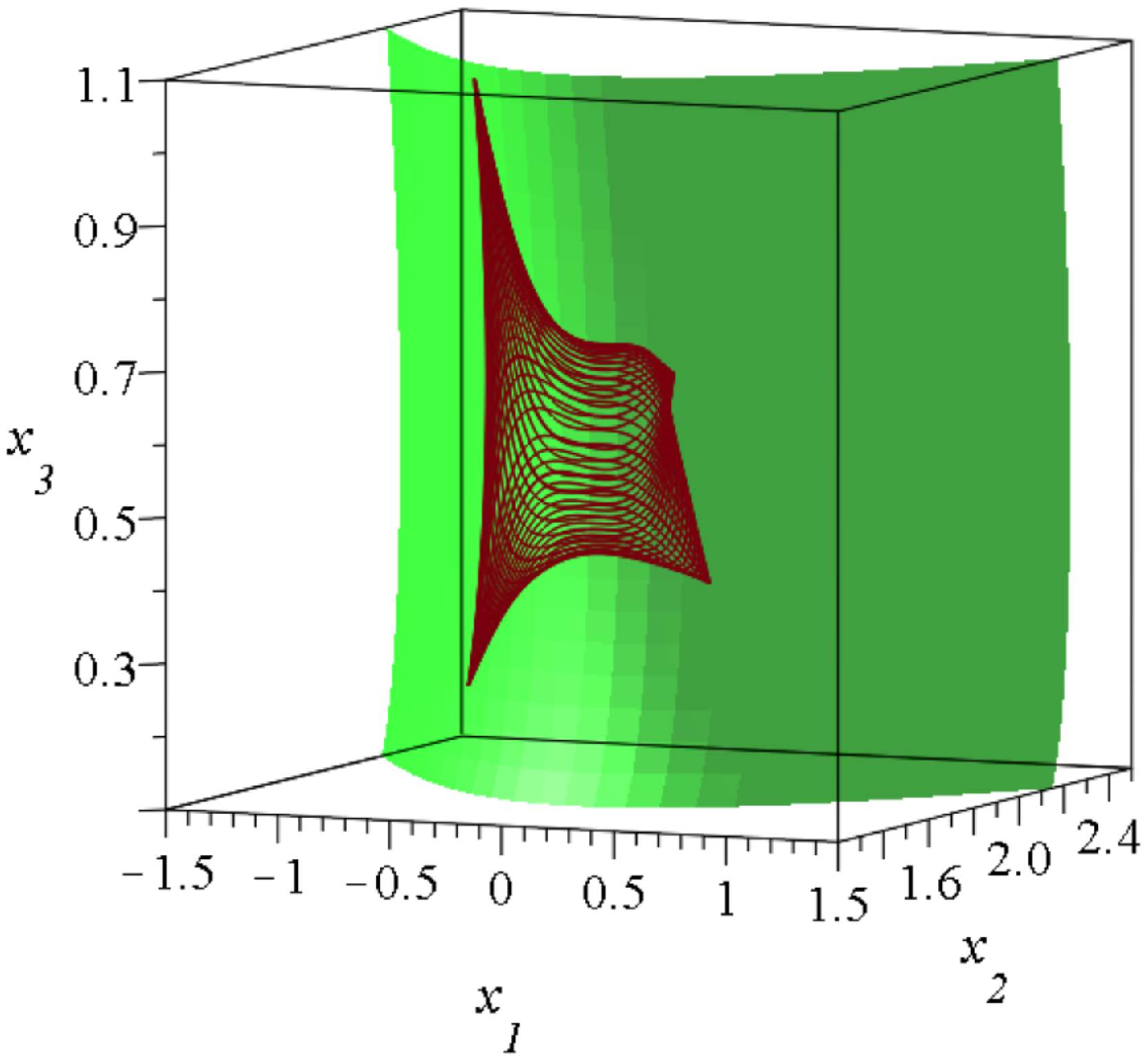}}
\subfloat[\label{integral_surfaces2}]
{\includegraphics[height=7.5cm, width=7.5cm]{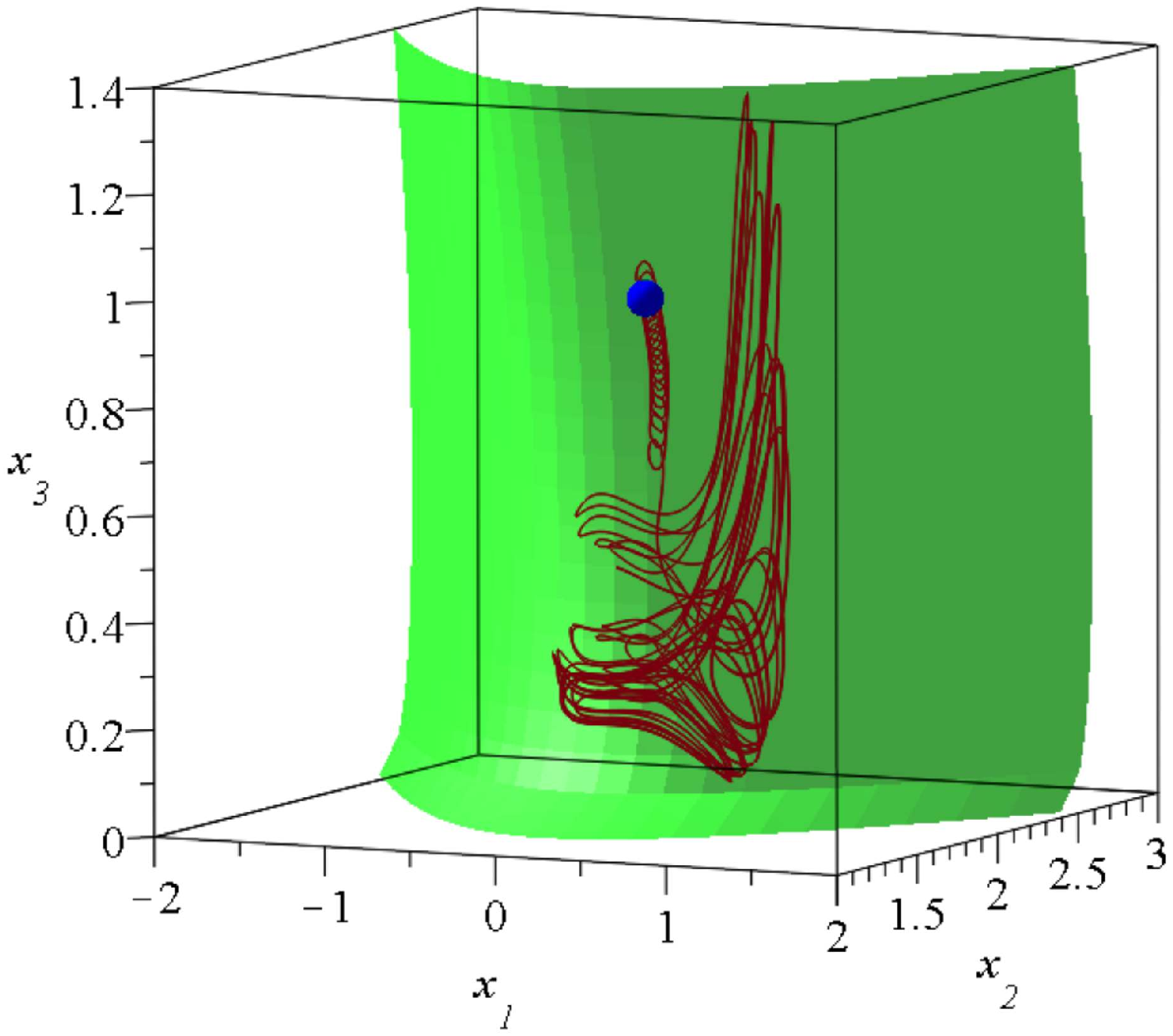}}
\caption{Two orbits in the case $\Psi(t)=\frac{1}{\sqrt{3}}\Big(\Psi_{0,0,0}(t)+\Psi_{1,1,0}(t)+\Psi_{1,0,2}(t)\Big)$. The first one is ordered  ($x_1(3)=0,x_2(3)=0.7,x_3(3)=1.6401$) while the second one is chaotic ($x_1(1.018576206) = .297, x_2(1.018576206) = 1.63,x_3(1.018576206)=1.05$). Both orbits lie on the surface $-x_1^2+x_2^2+\frac{x_3^2}{2}-\frac{\sqrt{3}\ln(|x_3|)}{6}=C$, with  $C=3.037948931$ ($\omega_1=1,\omega_2=\sqrt{2},\omega_3=\sqrt{3}$). The second orbit starts close to a nodal point of the wavefunction (blue dot). We observe the expected vortices around the nodal point in the first part of the orbit.}
\end{figure}

This surface is open, extending to infinity. In particular on this surface we see an nodal point (blue dot in  Fig.~\ref{integral_surfaces2}) and orbits starting close to it are in general chaotic, while orbits starting far from it are in general ordered (Fig.~\ref{integral_surfaces1}).

\begin{figure}[ht!]
\centering
\subfloat
{\includegraphics[height=6.5cm, width=7.5cm]{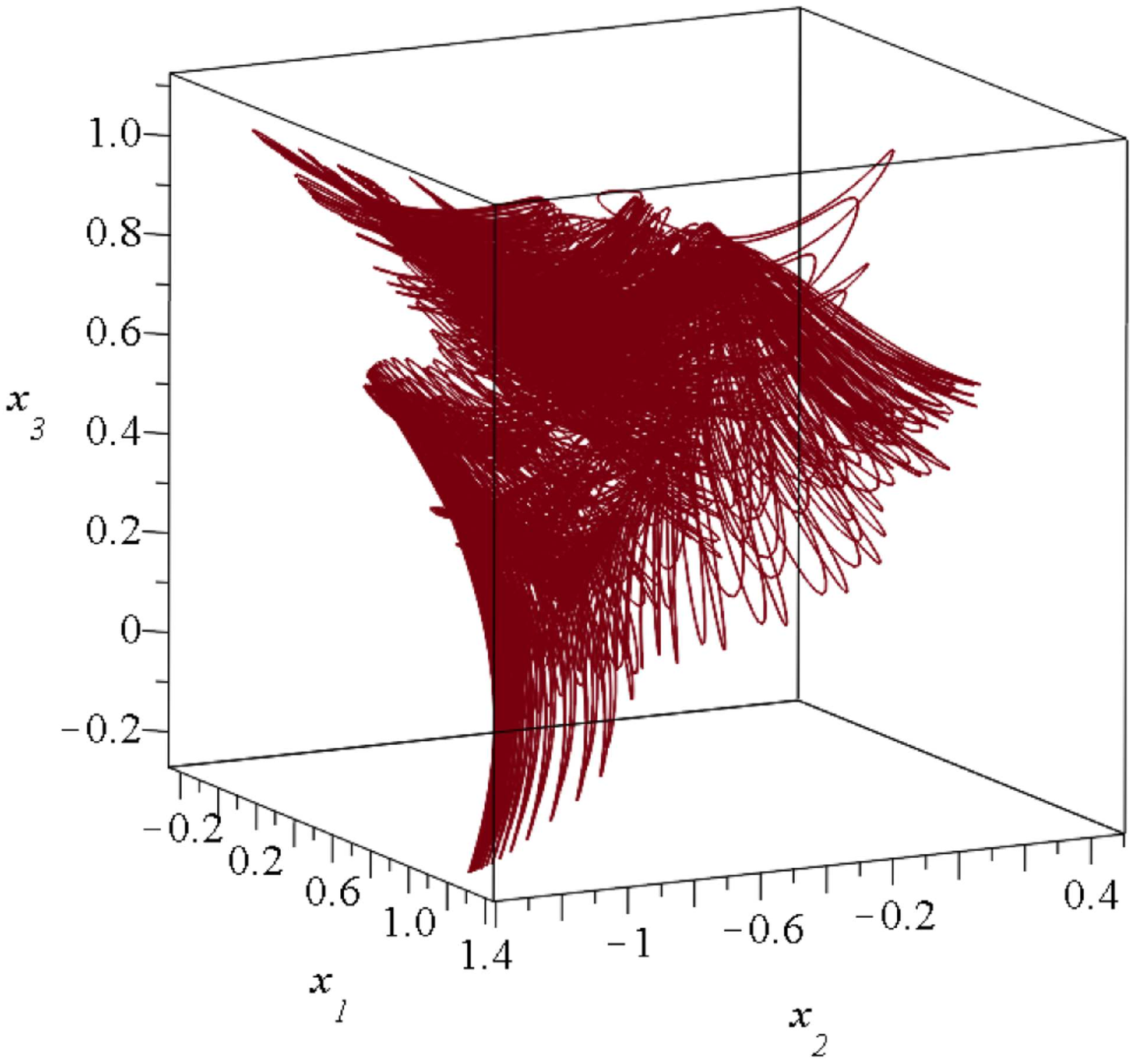}}
\subfloat
{\includegraphics[height=6.5cm, width=7.cm]{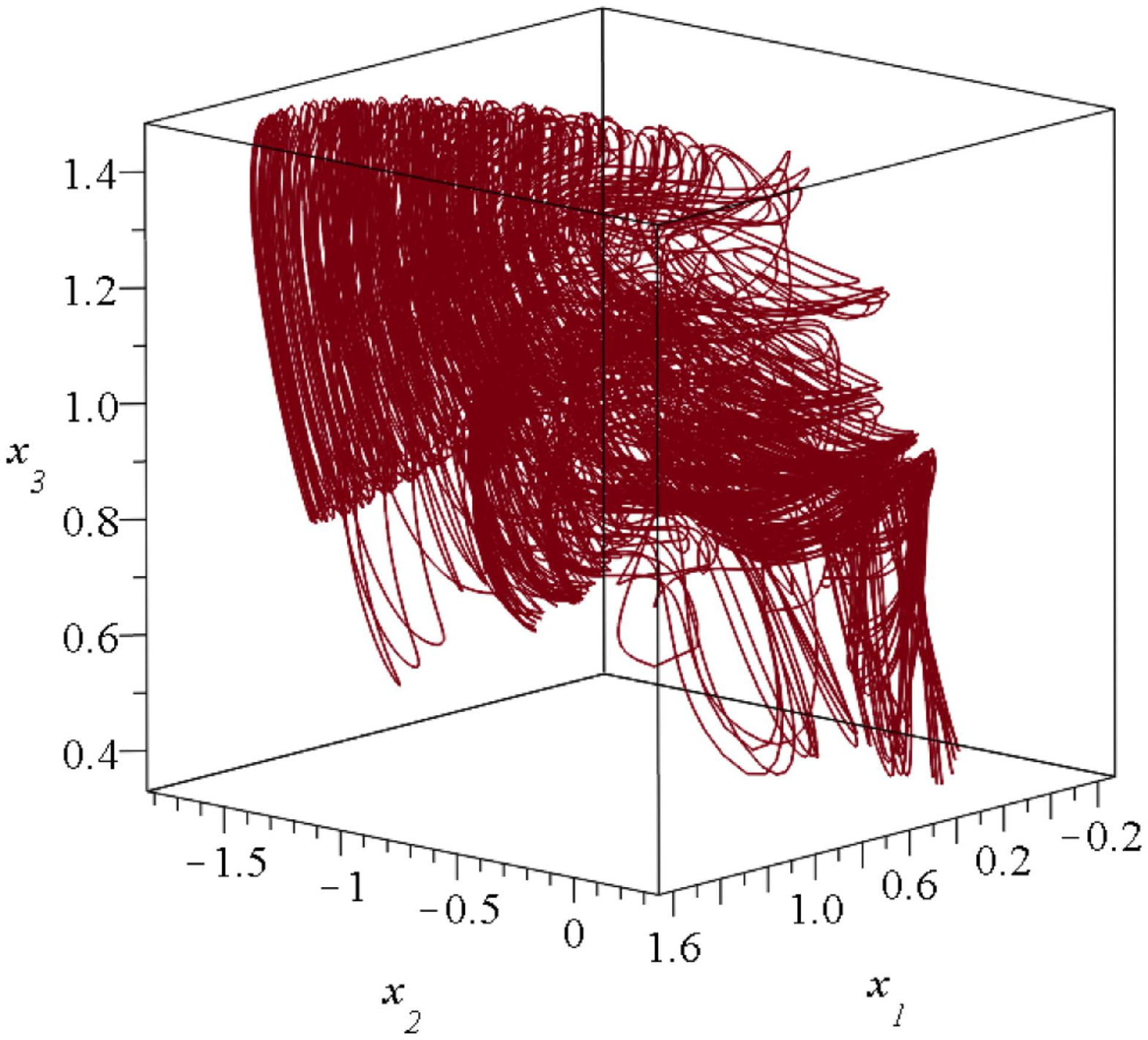}}
\caption{Two orbits in the case $\Psi(t)=\frac{1}{\sqrt{3}}\Big(\Psi_{0,0,0}(t)+\Psi_{1,1,0}(t)+\Psi_{0,2,1}(t)\Big)$ for different initial conditions ($x_1(0)=0.5,x_2(0)=0.5,x_3(0)=0.5$ and $x_1(0)=0.125,x_2(0)=0,x_3(0)=0.33$ correspondingly). Both of them are chaotic and do not lie on any invariant surface ($\omega_1=1,\omega_2=\sqrt{2},\omega_3=\sqrt{3}$).}\label{nods}
\end{figure}

}
\item{Case C \begin{align}\Psi(t)=\frac{1}{\sqrt{3}}\Big(\Psi_{0,0,0}(t)+\Psi_{1,1,0}(t)+\Psi_{0,2,1}(t)\Big)\end{align} 
This case does not belong to any of the cases 1-6 above. Therefore there is no integral  of the form (\ref{form}) in this case. Furthermore it is  quite improbable that there is any other form of an integral. In fact the orbits do not lie on any surface but are chaotic in 3-dimensions. Two such orbits are shown in Fig.~(\ref{nods}).

}
\end{itemize}

It is remarkable that although both cases B and C have the same indices $0,1,2$, differing only as regards the sequence of these numbers in the third term of the sum, $\Psi_{1,0,2}$ vs $\Psi_{0,2,1}$, their orbits are of a quite different character. In the first case the orbits lie on a smooth surface (both ordered and chaotic orbits), while in the second case they are all chaotic in 3 dimensions and do not lie on any surface.

\section{Connection with integrability in the 2-d and 4-d cases}
The results presented in the previous sections refer to particular choices of quantum states in 3-d systems of harmonic oscillators. We now show that similar arguments hold in the 2-d and 4-d cases. In fact, we conjecture that our method for finding partially integrable systems can be generalized to an arbitrary number of dimensions.

In the 2-d case we consider wavefunctions of the form
\begin{align}
\Psi(t)=a\Psi_{p1,p2}(t)+b\Psi_{r1,r2}(t)+c\Psi_{s1,s2}(t)
\end{align}
The corresponding equations of motion are:
\begin{align}\label{v1}
\nonumber&\dot{x}_1=\frac{1}{\tilde{G}}\Bigg(abK_1K_2H_{p_2}H_{r_2}[H_{p_1},H_{r_1}]\sin(\Delta E_{12}t)+acK_1K_3H_{p_2}H_{s_2}[H_{p_1},H_{s_1}]\sin(\Delta E_{13}t)\\&+bcK_2K_3H_{r_2}H_{s_2}[H_{r_1},H_{s_1}]\sin(\Delta E_{23}t)\Bigg)\\
\label{v2}&\nonumber\dot{x}_2=\frac{1}{\tilde{G}}\Bigg(abK_1K_2H_{p_1}H_{r_1}[H_{p_2},H_{r_2}]\sin(\Delta E_{12}t)+acK_1K_3H_{p_1}H_{s_1}[H_{p_2},H_{s_2}]\sin(\Delta E_{13}t)\\&+bcK_2K_3H_{r_1}H_{s_1}[H_{r2},H_{s_2}]\sin(\Delta E_{23}t)\Bigg)
\end{align}
where $K_j=\frac{(\omega_1\omega_2)^{\frac{1}{4}}}{\sqrt{\pi 2^{n_1+n_2}n_1!n_2!}},$  with $n_i=p_i$ if $j=1$, $n_i=r_i$ if $j=2$ and $n_i=s_i$ if $j=3$. Moreover $[H_l,H_m]=H_lH_m'-H_l'H_m$ and $\tilde{G}=e^{\omega_1x_1^2+\omega_2x_2^2}(\Psi_R^2+\Psi_I^2)$.

In order to find an integral we must eliminate 3 trigonometric terms, but we have only two equations. Thus in general we cannot find common factors to eliminate all three trigonometric terms. E.g. in order to eliminate the terms with $\sin(\Delta E_{23}t)$ we must take $r_1=s_1$ and then $[H_{p_1},H_{s_1}]=[H_{p_1},H_{r_1}]$. However, the corresponding term with $\sin(\Delta E_{32}t)$ in Eq.~(\ref{v2}) cannot be eliminated unless $r_2=s_2$. But then $E_2=E_3$ and the term $\Psi_{s_1,s_2}$ is not different from $\Psi_{r_1,r_2}$. Similarly any other atempt to eliminate all three trigonometric terms fails if $abc\neq 0$. A particular nonintegrable case of this form was considered in \cite{contopoulos2008ordered}, with $p_1=p_2=r_2=0$ and $r_1=s_1=1$ and $s_2=2$.

On the other hand, in a 2-d case with two terms (e.g. $c=0$) we have
\begin{align}
\dot{x}_1=\frac{1}{\tilde{G}}\Big(abK_1K_2H_{p_2}H_{r_2}[H_{p_1},H_{r_1}]\sin(\Delta E_{12}t)\Big)\\
\dot{x}_2=\frac{1}{\tilde{G}}\Big(abK_1K_2H_{p_1}H_{r_1}[H_{p_2},H_{r_2}]\sin(\Delta E_{12}t)\Big).
\end{align}
Therefore we can eliminate $\sin(\Delta E_{12}t)$ by multiplying $\dot{x}_1$ and $\dot{x}_2$ by appropriate functions so that their difference is zero
\begin{align}\label{diff}
\frac{\dot{x}_1H_{p_1}H_{r_1}}{[H_{r_1},H_{p_1}]}-\frac{\dot{x}_2H_{p_2}H_{r_2}}{[H_{r_2},H_{p_2}]}=0
\end{align}
thus we have an integral 
\begin{align}\label{int}
\int\frac{H_{p_1}H_{r_1}}{[H_{p_1},H_{r_1}]}dx_1-\int\frac{H_{p_2}H_{r_2}}{[H_{p_2},H_{r_2}]}dx_2=C,
\end{align}
provided that $p_1\neq r_1, p_2\neq r_2$ (If $p_i=r_i$ the corresponding term in Eqs.~(\ref{diff}, \ref{int}) is zero).
In these cases the system is completely integrable and there is no chaos.
Therefore if $\Psi$ is the sum of 2 terms we have an integral, but if $\Psi$ is the sum of 3 terms no integral exists. 
A special case is $\Psi(t)=\Psi_{1,0}(t)+\Psi_{0,1}(t)$. In this case the integral is: $x_1^2+x_2^2=C$ and the orbits are circles.

We conclude that if the wavefunction is the sum of a number greater than its dimension  there are no partially integrable cases that can be generated by our method, while such cases do appear if the number of eigenfunctions is equal to the dimension (with particular choices of the indices). In the case of $2$ dimensions a system of 2 terms is completely integrable but in the case of 3 dimensions a system of 3 terms is only partially integrable. Thus in the integrable 2-d case there is no chaos, while in partially integrable 3-d cases, both ordered and chaotic orbits appear on the same integral surface.

Finally, if in the 3-d case we have the sum of only two wavefunctions, then the system is completely integrable, i.e. it has two integrals of motion (special case ii), and no chaos appears.

Similar results are found in 4-d cases.
Then we have wavefunctions of the form
\begin{align}
\nonumber\Psi&=a\Psi_{p_1,p_2,p_3,p_4}(t)+b\Psi_{r_1,r_2,r_3,r_4}(t)\\&+c\Psi_{s_1,s_2,s_3,s_4}(t)+d\Psi_{t_1,t_2,t_3,t_4}(t)
\end{align}
and the equations of motion are:
\begin{align}\nonumber
\nonumber\dot{x}_1=&\frac{1}{\tilde{G}}\Big(abK_1K_2H_{p_2}H_{p_3}H_{p_4}H_{r_2}H_{r_3}H_{r_4}[H_{p_1},H_{r_1}]\sin(\Delta E_{12}t)\\&\nonumber+acK_1K_3H_{p_2}H_{p_3}H_{p_4}H_{s_2}H_{s_3}H_{s_4}[H_{p_1},H_{s_1}]\sin(\Delta E_{13}t)\\&\nonumber+adK_1K_4H_{p_2}H_{p_3}H_{p_4}H_{t_2}H_{t_3}H_{t_4}[H_{p_1},H_{t_1}]\sin(\Delta E_{14}t)\\&\nonumber+bcK_2K_3H_{r_2}H_{r_3}H_{r_4}H_{s_2}H_{s_3}H_{s_4}[H_{r_1},H_{s_1}]\sin(\Delta E_{23}t)\\&\nonumber+bdK_2K_4H_{r_2}H_{r_3}H_{r_4}H_{t_2}H_{t_3}H_{t_4}[H_{r_1},H_{t_1}]\sin(\Delta E_{24}t))\\&+cdK_3K_4H_{s_2}H_{s_3}H_{s_4}H_{t_2}H_{t_3}H_{t_4}[H_{s_1},H_{t_1}]\sin(\Delta E_{34}t)\Big)
\end{align}
and similar expressions for $\dot{x}_2, \dot{x}_3,\dot{x}_4$. In order to find an integral of motion we must eliminate the 6 trigonometric terms $\sin(\Delta E_{ij}t)$, $i=1\dots3, j=1\dots 4, i<j$. We cannot do this in general, but we can eliminate some trigonometric terms by appropriate selection of the indices $p,r,s,t$. E.g. by setting $p_1=r_1=s_1$ in $\dot{x}_1$ we eliminate the trigonometric terms $\sin(\Delta E_{12}t), \sin(\Delta E_{13}t)$ and $\sin(\Delta E_{23}t)$. Then we have also $[H_{p_1},H_{t_1}]=[H_{r_1},H_{t_1}]=[H_{s_1},H_{t_1}]$ and the appropriate factor of $\dot{x}_1$ is $H_{p_1}H_{t_1}/[H_{p_1},H_{t_1}]$. In a similar way we set $p_2=s_2=t_2$  in $\dot{x}_2$ and find the factor $f_2(x_2)=H_{p_2}H_{r_2}/[H_{p_2},H_{r_2}]$. We set also $r_3=s_3=t_3$ in $\dot{x}_3$ and find the factor $f_3(x_3)=H_{r_3}H_{p_3}/[H_{r_3},H_{p_3}]$ and $p_4=r_4=t_4$ in $\dot{x}_4$ and find $f_4(x_4)=H_{p_4}H_{s_4}/[H_{p_4},H_{s_4}]$. Then we find
\begin{align}\label{4d_c}
\frac{\dot{x}_1H_{p_1}H_{t_1}}{[H_{p_1},H_{t_1}]}+\frac{\dot{x}_2H_{p_2}H_{r_2}}{[H_{p_2},H_{r_2}]}+\frac{\dot{x}_3H_{p_3}H_{r_3}}{[H_{p_3},H_{r_3}]}+\frac{\dot{x}_4H_{p_4}H_{s_4}}{[H_{p_4},H_{s_4}]}=0
\end{align}
We check that this quantity is zero, because all trigonometric terms are cancelled. In fact every term $\dot{x}_if(x_i)$ contains now 3 trigonometric terms and all 4 terms contain 12 trigonometric terms, but every one of the original 6 trigonometric terms appears in the sum twice with opposite sign. As a consequence we find an integral of motion by integrating Eq.~(\ref{4d_c}).
Other combinations of the indices $p,r,s,t$ are also possible.

If $d=0$ we have only 3 terms in $\Psi$ and only three trigonometric terms $\sin(\Delta E_{12}t)$, $\sin(\Delta E_{13}t), \sin(\Delta E_{23}t)$ as in the case of 3-d systems. 

In this case we can ignore the 4th indices $p_4,r_4,s_4$ and the equation with $\dot{x}_4$ and find on integral of motion in the same way as in 3-d systems. E.g. we have a case similar to the 3-d case 1 if we set $p_1=r_1, r_2=s_2$ and $p_3=s_3$. Then $\dot{x}_1, \dot{x}_2, \dot{x}_3$ satisfy the relation
\begin{align}\label{app9}
\frac{\dot{x}_1H_{s_1}H_{p_1}}{[H_{s_1},H_{p_1}]}+\frac{\dot{x}_2H_{p_2}H_{r_2}}{[H_{p_2},H_{r_2}]}+\frac{\dot{x}_3H_{r_3}H_{s_3}}{[H_{r_3},H_{s_3}]}=0
\end{align}and its corresponding integral.
However in the 4-d case instead of ignoring the 4th indices we can ignore the first, or the second, or the third indices and we have similar relations. E.g. if  instead of the relation $p_3=s_3$ we put $p_4=s_4$, we have the relation
\begin{align}\label{app_10}
\frac{\dot{x}_1H_{s_1}H_{p_1}}{[H_{s_1},H_{p_1}]}+\frac{\dot{x}_2H_{p_2}H_{r_2}}{[H_{p_2},H_{r_2}]}+\frac{\dot{x}_4H_{r_4}H_{p_4}}{[H_{r_4},H_{p_4}]}=0
\end{align}
Therefore if we have four relations $p_1=r_1, r_2=s_2, p_3=s_3, r_4=s_4$ we have two integrals, by integrating Eqs.~(\ref{app9}) and (\ref{app_10}).  Then the orbits lie on 2-d surfaces which are the intersections of the integrals of Eqs.~(\ref{app9}) and (\ref{app_10}). If, however, we have the conditions $p_1=r_1, r_2=s_2, p_3=s_3$ but $p_4\neq s_4$ the orbits do not lie on 2-d surfaces, but they do belong to the 3-d integral surface found by integrating Eq.~(\ref{app9}).

Further combinations of the indices can be given as in the 3-d case.
Finally if $c=d=0$, the wavefunction $\Psi$ consists of only 2 terms and we have only one trigonometric term $\sin(\Delta E_{12}t)$. Then we find 3 relations between $\dot{x}_1,\dot{x}_2,\dot{x}_3,\dot{x}_4$, namely
\begin{align}
\frac{\dot{x}_1H_{p_1}H_{r_1}}{[H_{p_1},H_{r_1}]}=\frac{\dot{x}_2H_{p_2}H_{r_2}}{[H_{p_2},H_{r_2}]}=\frac{\dot{x}_3H_{p_3}H_{r_3}}{[H_{p_3},H_{r_3}]}=\frac{\dot{x}_4H_{p_4}H_{r_4}}{[H_{p_4},H_{r_4}]},
\end{align}
without any restrictions on the symbols $p_1, p_2, p_3, p_4, r_2, r_2, r_3, r_4$ assuming only that the terms $p_1, p_2,p_3, p_4$ are not all equal to $r_1, r_2, r_3, r_4$ (If some indices are equal the corresponding terms are zero). E.g . if $p_1=r_1$, we have $\dot{x}_1=0$ plus 2 more integrals involving $\dot{x}_2, \dot{x}_3,  \dot{x}_4$. Thus if $c=d=0$ there are 3 integrals of motion and all orbits are ordered as the system is reduced to 1 dimension and does not have chaos. On the other hand if $\Psi$ is the sum of more than 4 terms, then we cannot find combinations of the indices that would give an integral of motion of the form (\ref{form}).
\section{Conclusions}

Partial integrability of the Bohmian trajectories is a property of certain quantum systems with implications, both mathematical and physical. 
On the one hand it confines the trajectories, both ordered and chaotic, on certain invariant surfaces. Therefore the evolution of 3-d Bohmian trajectories is dictated by two (and not three) independent variables, something that facilitates the study of the emergence of Bohmian chaos by means of the nodal point- X-point mechanism \cite{Tzemos2016}.
On the other hand this confinement is responsible for the restriction of quantum relaxation. Consequently, the dynamical approximation of Born's rule by the  quantum probabilities cannot be accomplished in the presence of partial integrability.

In this paper we studied analytically the existence of partial integrability of 3-d Bohmian trajectories in a system composed of 3 harmonic oscillators with wavefunctions of the general form (\ref{general}). Our main results are:
\begin{enumerate}
\item{By inspection of the Bohmian equations of motion, we found that the proper choice of the quantum numbers is responsible for the existence or not of invariant surfaces. In particular, we presented an method that generates conserved quantities of the form $f(x_1)\dot{x}_1+f(x_2)\dot{x}_2+f(x_3)\dot{x}_3=0.$ By means of the above method, we presented all the possible partially integrable cases and their corresponding integrals of motion.}
\item{We gave some numerical examples of the above results. In particular we presented  two distinctive partially integrable systems: The first one is characterized by a closed surface, while the second one by an open surface. In both cases we showed the coexistence of order and chaos on the same surface.} 
\item{Then we presented one case in which there is no integral of motion of the form (\ref{form}), or of any other form. In fact in this case the orbits are completely chaotic. They are free to wander around the 3-d space, something necessary for the accomplishment of quantum relaxation and the dynamical approximation of Born's rule.}
\item{Finally, we made a technical discussion about 2-d and 4-d partially integrable cases  and compare them with the 3-d case. We show that the number of the terms of the wavefunction is important for the existence or not of an integral of motion by means of our proposed method. If we have more terms than the dimension of the system then partial integrability of the form (\ref{form}) does not exist. In general, we conjecture that our method for identifying partially integrable cases of Bohmian trajectories can be generalized to $N$ dimensions, with $N$ arbitrary.}
\end{enumerate}

\textbf{Acknowledgements}: This research is supported by the Research Commitee of the Academy of Athens.

\bibliographystyle{elsarticle-num}
\bibliography{Partial_Integrability_Revision_arxiv}

\begin{thebibliography}{10}
\expandafter\ifx\csname url\endcsname\relax
  \def\url#1{\texttt{#1}}\fi
\expandafter\ifx\csname urlprefix\endcsname\relax\def\urlprefix{URL }\fi
\expandafter\ifx\csname href\endcsname\relax
  \def\href#1#2{#2} \def\path#1{#1}\fi

\bibitem{debroglie1927a}
L.~De~Broglie, {C. R. Acad. Sci., Paris} 184 (1927a) 283.

\bibitem{debroglie1927b}
L.~De~Broglie, {C. R. Acad. Sci., Paris} 185 (1927b) 380.

\bibitem{Bohm}
D.~Bohm, A suggested interpretation of the quantum theory in terms of "hidden"
  variables. i, Phys. Rev. 85 (1952) 166.

\bibitem{BohmII}
D.~Bohm, A suggested interpretation of the quantum theory in terms of "hidden"
  variables. ii, Phys. Rev. 85 (1952) 180.

\bibitem{holland1995quantum}
P.~R. Holland, The quantum theory of motion: an account of the de Broglie-Bohm
  causal interpretation of quantum mechanics, Cambridge University Press, 1995.

\bibitem{deurr2009bohmian}
D.~D{\"u}rr, S.~Teufel, Bohmian Mechanics: The Physics and Mathematics of
  Quantum Theory, Springer, 2009.

\bibitem{benseny}
A.~Benseny, G.~Albareda, {\'A}.~S. Sanz, J.~Mompart, X.~Oriols, Applied
  {B}ohmian mechanics, Eur.Phys.J. D 68~(10) (2014) 1.

\bibitem{pladevall2012applied}
X.~O. Pladevall, J.~Mompart, Applied {B}ohmian mechanics: From nanoscale
  systems to cosmology, CRC Press, 2012.

\bibitem{frisk1997properties}
H.~Frisk, Properties of the trajectories in {B}ohmian mechanics, Phys. Lett. A
  227~(3) (1997) 139.

\bibitem{falsaperla2003motion}
P.~Falsaperla, G.~Fonte, On the motion of a single particle near a nodal line
  in the de broglie--bohm interpretation of quantum mechanics, Phys. Lett. A
  316~(6) (2003) 382.

\bibitem{wisniacki2005motion}
D.~A. Wisniacki, E.~R. Pujals, Motion of vortices implies chaos in {B}ohmian
  mechanics, Europhys Lett. 71~(2) (2005) 159.

\bibitem{efthymiopoulos2006chaos}
C.~Efthymiopoulos, G.~Contopoulos, Chaos in {B}ohmian quantum mechanics, J.
  Phys. A 39~(8) (2006) 1819.

\bibitem{wisniacki2007vortex}
D.~Wisniacki, E.~Pujals, F.~Borondo, Vortex dynamics and their interactions in
  quantum trajectories, J. Phys. A. 40~(48) (2007) 14353.

\bibitem{efthymiopoulos2007nodal}
C.~Efthymiopoulos, C.~Kalapotharakos, G.~Contopoulos, Nodal points and the
  transition from ordered to chaotic {B}ohmian trajectories, J. Phys. A 40~(43)
  (2007) 12945.

\bibitem{contopoulos2008ordered}
G.~Contopoulos, C.~Efthymiopoulos, Ordered and chaotic {B}ohmian trajectories,
  Celest. Mech. Dyn. Astron. 102~(1-3) (2008) 219.

\bibitem{PhysRevE.79.036203}
C.~Efthymiopoulos, C.~Kalapotharakos, G.~Contopoulos, Origin of chaos near
  critical points of quantum flow, Phys. Rev. E 79 (2009) 036203.

\bibitem{contopoulos2012order}
G.~Contopoulos, N.~Delis, C.~Efthymiopoulos, Order in de {B}roglie--{B}ohm
  quantum mechanics, J. Phys. A 45~(16) (2012) 165301.

\bibitem{Tzemos2016}
A.~C. Tzemos, G.~Contopoulos, C.~Efthymiopoulos, Origin of chaos in 3-d bohmian
  trajectories, Phys. Lett. A 380~(45) (2016) 3796.

\bibitem{VALENTINI19915}
A.~Valentini, Signal-locality, uncertainty, and the subquantum h-theorem. i,
  Phys. Lett. A 156~(1) (1991) 5.

\bibitem{VALENTINI19911}
A.~Valentini, Signal-locality, uncertainty, and the subquantum h-theorem. ii,
  Phys. Lett. A 158~(1) (1991) 1.

\bibitem{valentini2005dynamical}
A.~Valentini, H.~Westman, Proc. R. Soc. A 461~(2053) (2005) 253.

\end{thebibliography}

\end{document}